\documentclass[conference]{IEEEtran}
\IEEEoverridecommandlockouts
\usepackage{cite}
\usepackage{amsmath,amssymb,amsfonts}
\usepackage{algorithm}
\usepackage{algpseudocode}
\usepackage{graphicx}
\usepackage{psfrag}
\usepackage{subcaption}

\graphicspath{{./images/}}
\DeclareGraphicsExtensions{.pdf,.jpeg,.png}
\usepackage{textcomp}
\usepackage{xcolor}

\def\BibTeX{{\rm B\kern-.05em{\sc i\kern-.025em b}\kern-.08em
    T\kern-.1667em\lower.7ex\hbox{E}\kern-.125emX}}

\newcounter{tempEquationCounter}
\newcounter{thisEquationNumber}
\makeatletter
\newcommand\fs@spaceruled{\def\@fs@cfont{\bfseries}\let\@fs@capt\floatc@ruled
  \def\@fs@pre{\vspace{0.5\baselineskip}\hrule height.8pt depth0pt \kern2pt}%
  \def\@fs@post{\kern2pt\hrule\relax}%
  \def\@fs@mid{\kern2pt\hrule\kern2pt}%
  \let\@fs@iftopcapt\iftrue}
\makeatother

\newcommand{\bm}{\mathbf}
\newcommand{\be}{\begin{equation}}
\newcommand{\ee}{\end{equation}}
\newcommand{\bea}{\begin{eqnarray}}
\newcommand{\eea}{\end{eqnarray}}

\newcommand{\bG}{{\bf G}}

\begin{document}

\title{Comparison of OTFS and OFDM for RIS-aided Systems in the Presence of Phase Noise \\
}

\author{\IEEEauthorblockN{Stephen McWade, ~\IEEEmembership{Member, IEEE,}
 and 
Arman Farhang, ~\IEEEmembership{Senior Member, IEEE}}

\IEEEauthorblockA{Department of Electronic and Electrical Engineering, Trinity College Dublin, Ireland \\}

\IEEEauthorblockA{
Email: \{smcwade, arman.farhang\}@tcd.ie}

\thanks{This publication has emanated from research supported by Research Ireland under the US-Ireland R\&D Partnership Programme Grant Numbers 21/US/3757 and 24/US/4013.}}

\maketitle

\begin{abstract}
In this paper, we investigate the performance of RIS-aided orthogonal time frequency space (OTFS) and orthogonal frequency division multiplexing (OFDM) systems in the presence of oscillator phase noise. OFDM is known to be sensitive to phase noise, which could limit the potential gains promised by RIS systems. OTFS, on the other hand, is a compelling potential waveform for RIS-aided systems in the presence of phase noise due to it's resilience to time-varying channels. However, the effect of phase noise on OTFS has not been fully analyzed in the literature as of yet. Additionally, no existing works in the literature consider the effect of phase noise on an RIS-aided OTFS system. Hence, we propose a joint RIS channel and phase noise estimation technique using a Wiener filtering approach. Our proposed method exploits the statistical nature of both the phase noise and the Doppler spread channel in a setup with RIS. Our numerical analysis demonstrates the significant gain of RIS-aided OTFS offers compared to RIS-aided OFDM in the presence in the presence of  phase noise. Additionally, our results demonstrate the superiority of our proposed estimation technique, with gains of up to 3~dB in terms of bit error rate (BER), over existing methods in the literature.
\end{abstract}


\section{Introduction}
 Reconfigurable intelligent surface (RIS) are a rapidly developing concept in wireless communications technology in which a large number of passive reflecting elements with adjustable phase shifts to enhance network coverage, increase spectral efficiency and reduce energy consumption. \cite{DiRenzo_RIS}. In practical systems, RF impairments such as carrier frequency offset (CFO), in-phase (I) and quadrature-phase (Q) imbalances and oscillator phase noise are present and can negatively affect system performance \cite{RF_impairment_survey}. In the existing literature, orthogonal frequency division multiplexing (OFDM) has been widely considered as the modulation scheme for RIS-assisted systems \cite{Wu_IRS_tutorial_2021}. However, OFDM is known to be sensitive to phase noise, which leads to inter-carrier interference (ICI) and a significant degradation in bit error rate (BER) performance \cite{Fettweis_PN_OFDM}. Consequently, the combination of RIS with OFDM inherits this sensitivity, potentially negating the performance gains promised by RIS.

 Orthogonal Time Frequency Space (OTFS) modulation has been proposed as a more robust alternative to OFDM in scenarios involving high mobility and time-varying channels \cite{hadani_2017}. OTFS operates in the delay-Doppler domain and has been shown to offer superior performance over OFDM in rapidly changing environments, such as high-speed railway or vehicular communications. Importantly, phase noise is also a time-varying process which introduces a linear time-variant (LTV) behavior in the channel, for which OTFS is better suited \cite{gen_theory_PN}. This makes OTFS a compelling potential waveform for RIS-aided systems in the presence of phase noise. However, the effect of phase noise on OTFS has not been fully analyzed in the literature as of yet. In contrast to the channel taps in Doppler spread channels, which are generally considered as low-pass processes, phase noise is a wideband process. This means that phase noise and Doppler spread are fundamentally different in nature and channel estimation methods which work for the latter may not work for the former.

 As far as the authors are aware, there is no existing work in the literature which considers an RIS-aided OTFS system in the presence of phase noise. While several works have investigated RIS-aided OTFS schemes, they often assume ideal hardware and neglect the impact of phase noise \cite{Thomas_IRS_OTFS_2023}. Meanwhile, studies that consider conventional OTFS systems under phase noise typically consider perfect knowledge of phase noise at the receiver without offering detailed analysis \cite{Surabhi_OTFS_PN_2019, Bello_OTFS_pn}. Additionally, these works do not consider the effect of phase noise on an RIS-aided system.  This paper aims to bridge this gap by analyzing the performance of RIS-aided OTFS systems in the presence of phase noise and proposing a joint phase noise and channel estimation technique. 
 
 Hence, this paper addresses these gaps in the literature with the following contributions and results: (i) We study of the effect of phase noise on RIS-aided OTFS and RIS-OFDM systems. We present the structure of the phase noise matrix and its underlying effect on delay-Doppler multiplexing. (ii) We propose a joint RIS channel and phase noise estimation technique using  Wiener filtering to estimate the full effective channel response sample-by-sample. In this paper, we assume that the RIS reflection coefficients have already been designed at the base station using statistical channel information and thus our proposed technique is used to estimation the instantaneous effective channel comprise of the Doppler spread channel, phase noise and the RIS reflection coefficients. (iii) We present numerical results which compare the performance OTFS and OFDM in RIS-aided systems in the presence of phase noise. Our numerical analysis shows that OTFS provides up to two orders of magnitude performance gains, in terms of bit error rate (BER), over OFDM in RIS-aided systems the presence of phase noise. Additionally, we show that our proposed estimation technique provides BER gains of up to 3~dB compared to existing methods in the literature.

\subsubsection*{Notations} Superscripts ${(\cdot)^{\rm{T}}}$ and ${(\cdot)^{\rm{H}}}$ denote transpose and Hermitian transpose, respectively. Bold lower-case characters denote vectors and bold upper-case characters denote matrices. $x[n]$ denotes the $n$-th element of the vector $\mathbf{x}$. The matrix $\bm{F}_N$ is the $N$-point unitary discrete Fourier transform (DFT) matrix with $(l,k)$ elements $\frac{1}{\sqrt{N}}e^{-j\frac{2\pi}{N}lk}$ for $l,k=0,\ldots,N-1$. The functions $\mathrm{circ}\{ \mathbf{a} \}$ , and $\mathrm{diag}(\mathbf{a})$ form a circulant matrix whose first column is the vector $\mathbf{a}$ and a diagonal matrix with the diagonal elements in $\mathbf{a}$, respectively. The function $\rm{vec}(\mathbf{X})$ vectorizes the matrix $\mathbf{X}$ by concatenating its columns to form a vector. $\otimes$ represents the Kronecker product. The $p\times{p}$ identity matrix and $p \times q$ all-zero matrix are  denoted by $\mathbf{I}_p$ and $\mathbf{0}_{p\times{q}}$, respectively. Finally, $j = \sqrt{-1}$ represents the imaginary unit.
\section{System Model}
In this section, we present the system model of both RIS-assisted OFDM and RIS-assisted OTFS in the presence of phase noise. To begin we consider a generic RIS-aided system in the time domain.  We consider an RIS-aided system in which the RIS contains $Q$ reflecting elements whose phases can be tuned to steer the signal towards the receiver. In this paper, we consider a scenario where the direct line of sight path is impeded. Let $s(t)$ be the time domain transmit signal. After modulating the transmit signal to the carrier frequency, it propagates through a linear time-varying (LTV) channel to each RIS element. Hence, the continuous-time received signal in baseband at RIS element $i$ is represented as 
\begin{equation}
    z_i(t) = \int \int u^{i}(\tau_\mathrm{u},\nu_\mathrm{u})s(t-\tau_\mathrm{u})e^{j2\pi\nu_\mathrm{u}(t-\tau_\mathrm{u})}d\tau_\mathrm{u} d\nu_\mathrm{u}, 
\end{equation}
where $u^{i}(\tau_\mathrm{u},\nu_\mathrm{u}) = \sum_{p=0}^{P_\mathrm{u}-1}u^i_{p}\delta(\tau - \tau^i_{u,p})\delta(\nu - \nu^i_{u,p}),$ is the delay-Doppler domain channel impulse response (CIR), which consists of $P_\mathrm{u}$ channel paths. The parameters $u^i_{p}$, $\tau^i_{u,p}$ and $\nu^i_{u,p}$ represent the channel gain, delay and Doppler shift associated with path $p$ of the $i$-th TX-to-RIS element channel,  respectively. After reflecting off the RIS element with reflection coefficient $\varphi_i$, the signal travels through another LTV channel from the RIS to the receiver. In practical systems, local oscillator imperfections induce unwanted phase noise into the signal at both the transmitter and receiver sides. However, when the phase noise bandwidth at the transmitter and receiver side is small relative to the subcarrier spacing, the resulting phase noise effect is equivalent to a single phase noise at the receiver with a bandwidth equal to the sum of the bandwidth of the transmitter and receiver side processes \cite{Fettweis_PN_OFDM}. Thus, in this paper, we consider a system where the signal is affected by phase noise at the receiver side only. Hence, the received signal is given by
\begin{equation}
    r_i(t) = \varphi^ie^{j\theta(t)}\int \int v^{i}(\tau_\mathrm{v},\nu_\mathrm{v})z_i(t-\tau_\mathrm{v})e^{j2\pi\nu_\mathrm{v}(t-\tau_\mathrm{v})}d\tau_\mathrm{v} d\nu_\mathrm{v}, 
\end{equation}
where $v^{i}(\tau_\mathrm{v},\nu_\mathrm{v}) = \sum_{p=0}^{P_\mathrm{v}-1}v^i_{p}\delta(\tau - \tau^i_{v,p})\delta(\nu - \nu^i_{v,p}),$ is the delay-Doppler domain channel impulse response (CIR), which consists of $P_\mathrm{v}$ channel paths. The parameters $v^{i}_{p}$, $\tau^i_{v,p}$ and $\nu^i_{v,p}$ represent the channel gain, delay and Doppler shift associated with path $p$ of the $i$-th RIS-to-RX channel,  respectively. The term $\theta(t)$ represents the continuous-time phase noise angle at time $t$.

The received signal then sampled with sampling period $T_{\rm{s}}$. The sampling period is assumed to be short, such that the path delays can be considered as integer multiples of the sampling period. However, the Doppler shifts cannot be considered to be integer multiples of the Doppler spacing and thus, we consider fractional Doppler shifts in this paper. The discrete received signal samples can then be expressed as The discrete-time received signal for the $i$-th RIS element can be expressed as  
 \begin{equation}
 \begin{split}
     r_i[n] &= \varphi_ie^{j\theta[n]}\sum_{l_\mathrm{v}=0}^{L_\mathrm{v}-1}v_i[n,l_\mathrm{v}]z_i[n-l_\mathrm{v}],
     \\& = \varphi_ie^{j\theta[n]}\sum_{l_\mathrm{v}=0}^{L_\mathrm{v}-1}v_i[n,l_\mathrm{v}]\sum_{l_\mathrm{u}=0}^{L_\mathrm{u}-1}u_i[n,l_\mathrm{u}]s[n-l_\mathrm{v} - l_\mathrm{u}],
 \end{split}
       \label{eq:7}
\end{equation}
where $z_i[n]$, $u_i[n]$, $v_i[n]$, $\theta[n]$ and $s[n]$ are the discrete-time baseband equivalents of $z_i(t)$, $u_i(t)$, $v_i(t)$, $\theta(t)$ and $s(t)$, respectively.
By stacking the received signal samples into an $MN \times 1$ vector, (\ref{eq:7}) can be represented in matrix form as
\begin{equation}
    \mathbf{r}_i = \varphi_i\mathbf{\Phi}_{\rm{DT}}\mathbf{H}_{\mathrm{DT},i}\mathbf{s},
\end{equation}
The receiver receives a summation of all the signals reflected off each RIS element and thus, the resulting received signal vector is given by
\begin{equation}
    \mathbf{r} = \mathbf{\Phi}_{\rm{DT}}\mathbf{H}^{\mathrm{ris}}_\mathrm{DT}\mathbf{s} + \boldsymbol{\eta}
\end{equation}
where $\mathbf{H}^{\mathrm{ris}}_\mathrm{DT} = \sum_{i=0}^{Q}\varphi_i\mathbf{H}_{\mathrm{DT},i}$ is the full RIS delay-time domain channel matrix of size $MN \times MN$, $\mathbf{\Phi}_{\rm{DT}} = \mathrm{diag}(\boldsymbol{\psi}) $ is the phase noise matrix in the delay-time domain and 
\begin{equation}
    \boldsymbol{\psi} = [e^{j\theta[0]}, e^{j\theta[1]},\ \dots \ , e^{j\theta[MN - 1 ]}]^{\rm{T}},
\end{equation}
is the vector containing the phase noise at each sampling point. Additionally, $\boldsymbol{\eta}$ is the vector of complex additive white Gaussian noise (AWGN) with variance $\sigma_{\eta}^2$. In the following subsections we present the specific system model for RIS-aided OTFS and OFDM, respectively.

\subsection{RIS-assisted OTFS}
We consider an RIS-aided OTFS system $M$ Doppler bins and $N$ delay bins with delay and Doppler spacings of $\Delta\tau$ and $\Delta\nu$, respectively \cite{RCP_OTFS_Rav}. Let the $M \times N$ matrix $\mathbf{X}_{\mathrm{otfs}}$ contain the transmit quadrature amplitude modulation (QAM) data symbols on its elements in the delay-Doppler domain. The data symbols are assumed to be independent and identically distributed (i.i.d.) complex random variables of unit average power. The delay-time transmit signal is obtained by taking an $N$-point IDFT across the rows of $\mathbf{X}_{\mathrm{otfs}}$, i.e., the Doppler dimension \cite{Farhang_letter}. The signal then undergoes parallel to serial conversion and a cyclic prefix of length $N_{\rm{cp}}$ is appended to the beginning of the transmit signal. Thus, the time domain OTFS transmit signal is given by
\begin{equation}
    \mathbf{s}_{\mathrm{otfs}} = \mathbf{A}_{\mathrm{cp}}(\mathbf{F}_{N}^{\rm{H}}\otimes\mathbf{I}_{M})\mathbf{x}_{\mathrm{otfs}},
\end{equation} 
where $\bm{x} = \rm{vec}(\bm{X}_{\mathrm{otfs}})$. By substituting (7) into (5), the received signal is given by
\begin{equation}
    \mathbf{r} = \mathbf{\Phi}_{\rm{DT}}\mathbf{H}^{\mathrm{ris}}_\mathrm{DT}\mathbf{s}_{\mathrm{otfs}} + \boldsymbol{\eta}
\end{equation}
The received signal is then converted back to the delay-Doppler domain by a DFT operation along the time dimension, i.e.,
\begin{equation}
\begin{split}
    \mathbf{y}_{\mathrm{otfs}} &=(\mathbf{F}_{N}\otimes\mathbf{I}_{M})\mathbf{r} + (\mathbf{F}_{N}\otimes\mathbf{I}_{M})\boldsymbol{\eta} \\
    &=\mathbf{\Phi}_{\rm{DD}}\mathbf{H}^{\mathrm{ris}}_\mathrm{DD}\mathbf{x}_{\mathrm{otfs}} + \mathbf{w}_{\mathrm{dd}}.
    \end{split}
\end{equation}
The delay-Doppler domain phase noise matrix is given by $\mathbf{\Phi}_{\rm{DD}} = (\mathbf{F}_{N}\otimes\mathbf{I}_{M})\mathbf{\Phi}_{\rm{DT}}(\mathbf{F}_{N}^{\rm{H}}\otimes\mathbf{I}_{M})$ and the delay-Doppler domain representation of the LTV channel is given by $\mathbf{H}_{\mathrm{DD}} = (\mathbf{F}_{N}\otimes\mathbf{I}_{M})\mathbf{H}_{\mathrm{DT}}(\mathbf{F}_{N}^{\rm{H}}\otimes\mathbf{I}_{M}).$  The matrix $\mathbf{\Phi}_{\rm{DD}}$ has a block circulant, i.e.,

\begin{equation}
    \mathbf{\Phi}_{\rm{DD}} = \mathrm{circ}\{ [\mathbf{\Psi}_0^{\mathrm{T}},\ \dots \ , \mathbf{\Psi}_{N-1}^{\mathrm{T}}]^{\mathrm{T}} \} \label{eq9}
\end{equation}
where each $M\times M$ block matrix $\mathbf{\Psi}_n$ has a diagonal structure and is given by
$\mathbf{\Psi}_{n} = \mathrm{diag}([\phi_n[0], \phi_n[1],\ \dots \ , \phi_{n}[M-1]]^{\rm{T}}).$
The delay-Doppler domain phase noise coefficients which make up the elements of $\mathbf{\Phi}_{\rm{DD}}$ are each given by
\begin{equation}
    \phi_n[m] = \frac{1}{\sqrt{N}}\sum_{i=0}^{N-1}e^{j\theta[m + nM]}e^{-j\pi n i/N}.
\end{equation}

\subsection{RIS-assisted OFDM}
We consider an OFDM system with $M$ subcarriers and $N^{'}$ time domain symbols with subcarrier spacing $\Delta f$ and symbol duration $T = MT_{\rm{s}}$, respectively. We choose $N^{'} = \lceil \frac{MN+N_{\rm{cp}}}{M+N_{\rm{cp}}  + 2L-1} \rceil$, to maintain a OFDM frame length equal to the OTFS frame length after taking the CP and pilot insertion into account, where $L = L_v +L_u -1$ is the length of the RIS cascade channel. The delay-time transmit signal is obtained by taking an $M$-point IDFT across the columns of $\mathbf{X}_{\mathrm{ofdm}}$. The signal then undergoes parallel to serial conversion and therefore, the delay-time domain transmit signal can be written in vectorized form as
\begin{equation}
    \mathbf{s}_{\mathrm{ofdm}} = (\mathbf{I}_{N^{'}}\otimes\mathbf{A}_{\mathrm{cp}}\mathbf{F}_{M}^{\rm{H}})\mathbf{x}_{\mathrm{ofdm}}
\end{equation} 
The signal is then transmitted through the  channel. The received OFDM signal, considering the RIS and the phase noise effect is given by
\begin{equation}
    \mathbf{r}_{\mathrm{ofdm}} = \mathbf{\Phi}_{\rm{DT}}\mathbf{H}^{\mathrm{ris}}_\mathrm{DT}\mathbf{s}_{\mathrm{ofdm}} + \boldsymbol{\eta}.
\end{equation}
The cyclic prefix is then removed from the beginning of each OFDM symbol and an $M$-point DFT is applied to each individual OFDM symbol to convert the signal back to the frequency domain. The vectorized form of the overall received signal frame is given by
\begin{equation}
\begin{split}
    \mathbf{y}_{\mathrm{ofdm}} &= (\mathbf{I}_{N^{'}}\otimes\mathbf{F}_{M}\mathbf{R}_{\mathrm{cp}})\mathbf{r}_{\mathrm{ofdm}} + (\mathbf{I}_{N^{'}}\otimes\mathbf{F}_{M}\mathbf{R}_{\mathrm{cp}})\boldsymbol{\eta}
    \\& = \mathbf{\Phi}_{\rm{tf}}\mathbf{H}^{\mathrm{ris}}_\mathrm{tf} + \mathbf{w}_{\mathrm{tf}}
    \end{split}
\end{equation}
where the $\mathbf{R}_{\mathrm{cp}} = [\mathbf{0}_{M\times N_{\rm{cp}}},\mathbf{I}_M]$ is the cyclic prefix removal matrix. The time-frequency domain phase noise matrix is given by $\mathbf{\Phi}_{\rm{tf}} = (\mathbf{I}_{N^{'}}\otimes\mathbf{F}_{M}\mathbf{R}_{\mathrm{cp}})\mathbf{\Phi}_{\rm{DT}}(\mathbf{I}_{N^{'}}\otimes\mathbf{A}_{\mathrm{cp}}\mathbf{F}_{M}^{\rm{H}})$ and the time-frequency domain representation of the LTV channel is given by $\mathbf{H}_{\mathrm{tf}} = (\mathbf{I}_{N^{'}}\otimes\mathbf{F}_{M}\mathbf{R}_{\mathrm{cp}})\mathbf{H}_{\mathrm{DT}}(\mathbf{I}_{N^{'}}\otimes\mathbf{A}_{\mathrm{cp}}\mathbf{F}_{M}^{\rm{H}}).$ 
\section{Proposed RIS Channel and Phase Noise Estimation Technique}
In this section, we propose a joint phase noise and channel estimation technique for RIS-aided OTFS and OFDM systems. In time-varying channels it is extremely difficult to optimise the RIS reflection coefficients based on instantaneous channel information. Hence, in this paper, we assume that the RIS reflection coefficients have already been designed at the base station prior to the the transmission frame. Our proposed technique is then used to estimate the instantaneous effective channel, which includes phase noise, Doppler spread channel and the reflection coefficients, during the data transmission phase. The proposed technique has 2 stages. In the first stage, the well known impulse pilot in the delay-time domain is used to estimate the channel at the pilot indices \cite{Raviteja_OTFS_2019}. We utilise the pilot patterns for OTFS and OFDM systems from \cite{Huang_OTFS_OFDM} which can be seen in Fig. \ref{fig:impulse2}. The second stage uses Wiener filtering, based on the statistical nature of the phase noise and Doppler spread processes, to estimate the missing channel samples between the delay-time impulse pilot estimates.

\subsection*{Stage 1: Partial channel estimation}

\begin{figure}[t]
    \begin{subfigure}{\columnwidth}
        \includegraphics[width=\textwidth]{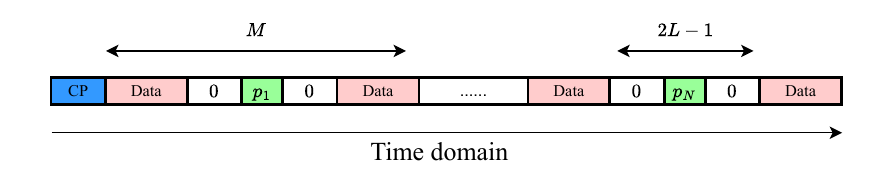}
        \caption{OTFS signal.}
    \end{subfigure}

    \begin{subfigure}{\columnwidth}
        \includegraphics[width=\textwidth]{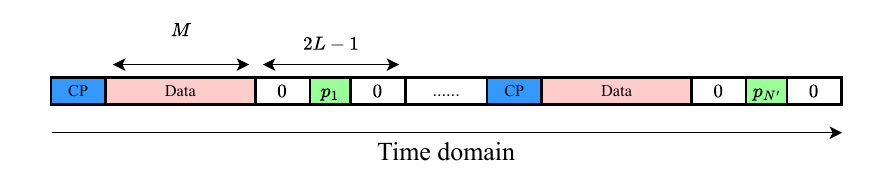}
        \caption{OFDM signal.}
    \end{subfigure}
    \caption{Illustration of the delay-time domain pilot patterns used for obtaining the initial estimate of the channel and phase noise for OTFS and OFDM systems.}\label{fig:impulse2}
\end{figure}

 The set of delay-time pilot indices is given by $\mathcal{P} = \{{m_1}, {m_2}, \ \dots \ ,{m_N}\}$. This pilot structure in the delay-time domain is illustrated in Fig. \ref{fig:impulse2}. We then use a threshold based method to estimate each of the channel delay taps at the pilot indices \cite{Raviteja_OTFS_2019}. The delay-time channel impulse response is estimated at the pilot region by capturing the $L \times N$  channel impulse response matrix at the delay bin at indices $\{m_p, m_p +1,\ \dots \ ,m_p +L-1\}$. The estimated channel impulse response matrix is which is given by
\begin{equation}
    \widehat{\mathbf{G}}_{\mathrm{DT}}^{\mathcal{P}} = [\widehat{\mathbf{g}}_{\rm{0}}, \widehat{\mathbf{g}}_{\rm{1}},\ \dots \ ,\widehat{\mathbf{g}}_{{L-1}}]^{\rm{T}},
\end{equation}
 where $\widehat{\mathbf{g}}_{{l}}$ is a $N\times1$ vector containing the channel estimates of the $l$-th channel delay tap at each of the pilot indices in $\mathcal{P}$. This stage only obtains the channel estimate at pilot location on the delay-time grid. However, as the actual channel is varying from sample to sample due to the phase noise and Doppler spread, there are missing channel samples in between the pilots. Therefore, at the next stage, we propose a statistical approach to obtain the sample-to-sample channel estimate, without the need for an increase in pilot overhead.

\subsection*{Stage 2: Full Channel Estimation} 
In the second stage, we propose a Wiener filter channel estimation technique based on the statistical properties of both the Doppler spread channel and the phase noise to improve upon the partial channel estimate from stage 1. The aim of our proposed channel estimation technique is to design a filter whose weights are captured in the matrix $\mathbf{W}$ which minimizes the mean-squared error between the estimated channel and actual effective channel for each of the channel taps. Hence, this problem can be written as 
\begin{equation}     
            \min_{\mathbf{W}}    \qquad \mathbb{E}[|\mathbf{W}\widehat{\mathbf{g}}_{\rm{l}} - \mathbf{g}_l|^2],\label{W_min}
\end{equation}
where $\mathbf{g} = [ \psi[0]h[0,l], ... ,\psi[MN-1]h[MN-1,l]]^{\rm{T}}$ is the vector containing the effective channel samples of channel tap $l$. Since each delay tap experiences the same phase noise and Doppler spread statistics, (\ref{W_min}) leads to the same solution for all values of $l$ and the only difference between the taps is the tap gain which scales the whole equation. Thus, we only need to solve (\ref{W_min}) for one delay tap and can apply the solution to each column to each column of $\widehat{\mathbf{G}}_{\mathrm{DT}}^{\mathcal{P}}$ as they are already scaled by the tap gains. Hence, without loss of generality, we we omit the subscript $l$ is the following equations.  
The solution to (\ref{W_min}) can be obtained via solving the Wiener-Hopf equations \cite{Adaptive_filters_book} and is given by 
$
    \mathbf{W} = \boldsymbol{K}_{g,\widehat{g}}(\boldsymbol{K}_{\widehat{g},\widehat{g}})^{\dag},
$
where  the $MN\times N$ $\boldsymbol{K}_{g,\widehat{g}} = \mathbb{E}\left[\mathbf{g}\widehat{\mathbf{g}}^{\rm{H}} \right]$ is the cross-correlation matrix of the effective channel and the estimated channel at the pilot indices. The $N\times N$ matrix $\boldsymbol{K}_{\widehat{g},\widehat{g}} = \mathbb{E}\left[\widehat{\mathbf{g}}\widehat{\mathbf{g}}^{\mathrm{H}} \right]$ is the autocorrelation matrix of the estimated channel at the pilot indices and is given by
\begin{equation}
\begin{split}
    \boldsymbol{K}_{\widehat{g},\widehat{g}} &= \mathbb{E}\left[\widehat{\mathbf{g}}\widehat{\mathbf{g}}^{\rm{H}} \right]\\
    &= \left(\boldsymbol{K}_{g,g}^\mathcal{P} + \frac{\sigma_\eta^2}{\sigma_{\mathrm{p}}^2}\mathbf{I}_N\right).
\end{split} \label{K_gg}   
\end{equation}
In (\ref{K_gg}), $\boldsymbol{K}_{g,g}^\mathcal{P}$ is the effective channel autocorrelation matrix at the pilot indices and $\sigma_{\mathrm{p}}^2$ represents the pilot power. The elements of $\boldsymbol{K}_{g,\widehat{g}}$ and $\boldsymbol{K}_{g,g}^\mathcal{P}$ are obtained by 
${K}_{g,\widehat{g}}[m,n] = {K}_{g}[m,p_n],$
 and 
 ${K}_{g,g}^\mathcal{P}[m,n] = {K}_{g}[p_m,p_n],$
 respectively, where ${K}_{g}[p_m,p_n]$ is the element $[p_m, p_n]$ of $\boldsymbol{K}_\mathrm{g}$, autocorrelation matrix of the full effective channel.
 
Since the Doppler spread and the oscillator phase noise are statistically independent, $\boldsymbol{K}_\mathrm{g}$ can be obtained via the product of the autocorrelation of matrices of the phase noise and the Doppler spread channel,
$\boldsymbol{K}_\mathrm{g} =  \boldsymbol{K}_{\psi}\boldsymbol{K}_\mathrm{D},$
where $\boldsymbol{K}_\mathrm{D}$ is the autocorrelation matrix of the Doppler spread channel and $\boldsymbol{K}_{\psi}= \mathbb{E}\left[\boldsymbol{\psi}\boldsymbol{\psi}^{\rm{H}} \right]$ is the autocorrelation matrix of the phase noise process. Assuming the Doppler effect follows the well-known Jakes model \cite{jakes_model}, the elements of $\boldsymbol{K}_\mathrm{D}$ are given by 
\begin{equation}
    {K}_\mathrm{D}[m,n] =  J_0(2\pi f_\mathrm{D}T_{\rm{s}}|m-n|)
\end{equation}
where $J_0(\cdot)$ is the zeroth order Bessel function of the first kind and $f_\mathrm{D}$ is the maximum Doppler shift in the channel. 

For the phase noise autocorrelation, the elements of $\boldsymbol{K}_{\psi}$ are given by
\begin{equation}
\begin{split}
    {K}_{\psi}[m,n] &= \mathbb{E}\left[ \psi[m]\psi^{*}[n]\right] \\
    &= \mathbb{E}\left[e^{j\Delta\theta[{(m-n)}]}\right],\\
\end{split} 
\label{K_psi}
\end{equation}

Since phase noise is a wide-sense stationary process, the expected value term is given by \cite{Fettweis_PN_OFDM}
\begin{equation}
    \mathbb{E}\left[e^{j\Delta\theta[{\delta}]}\right] = e^{-\frac{\sigma^2_{\theta}(\delta)}{2}},
\end{equation}
where $\sigma^2_{\theta}(\delta)$ is the time dependent variance or variogram of the phase-noise process \cite{Chorti_PN_spectral_model}.
The enumeration of the variogram depends on the type of oscillator being used by the transmitter and receiver. Thus, for the sake of completeness, we consider two oscillator types, namely,  free-running oscillator and continuous-time PLL in the following subsections.

\subsubsection{Free-Running Oscillator}
The simplest type of oscillator we can consider is a free-running oscillator. In this scenario, the phase noise is modeled as a Wiener process where the phase noise at time $n$ is given by
\begin{equation}
    \theta[n] = \theta[n-1] + \epsilon[n].
\end{equation}
The phase error at time $n$ is modeled as
$ \epsilon[n] \sim \mathcal{N}(0, \nu^2_{\mathrm{pn}}),  $
and $\nu^2_{\mathrm{pn}}$ is the sample-to-sample phase noise variance. For a free-running oscillator the phase noise variance is given by
$\nu^2_{\mathrm{pn}} = 4\pi\beta_{\mathrm{pn}} T_{\rm{s}}, $
where $\beta_{\mathrm{pn}}$ is the one-sided 3~dB line width of the oscillator's Lorentzian power spectral density.
In this scenario, the delay-dependent variance is given by
\begin{equation}
    \sigma^2_{\theta}(\delta) = 4\pi\beta_{\mathrm{pn}} T_{\rm{s}} \delta.
\end{equation}

\subsubsection{Continuous-time Phase Locked Loop}

\begin{figure}[t]
    \centering
\includegraphics[width=\columnwidth]{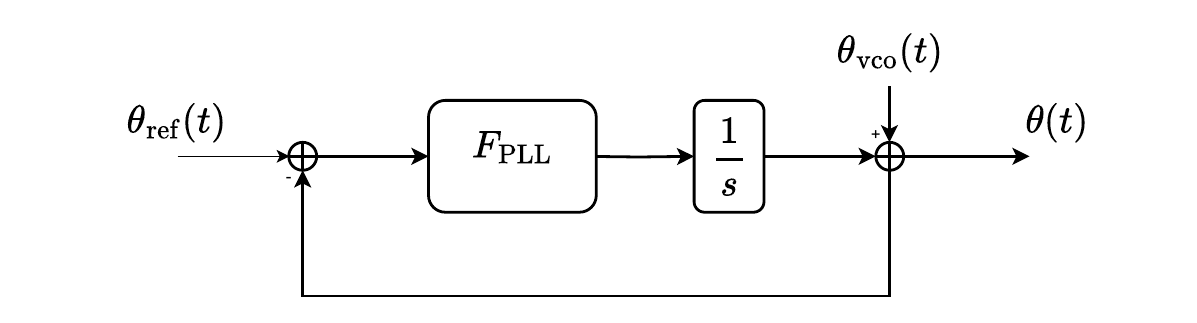}
    \caption{ Block diagram illustration of a first order continuous-time PLL diagram }
    \label{fig:contPLL_fig}
\end{figure}

In practical systems a PLL control system is used to generate a more stable oscillator output signal than a simple  free-running oscillator. In this paper, we consider a first order continuous-time PLL, see the block diagram in Fig \ref{fig:contPLL_fig}. The system consists of a reference signal, a VCO and a filter with filter coefficient $F_{\rm{PLL}}$. For this work, we assume a noiseless reference oscillator and noisy VCO oscillator. This output phase noise is modeled as an Ornstein-Uhlenbeck process as opposed to a Wiener process \cite{Fettweis_PN_OFDM}. Assuming Wiener noise  with zero mean and variance $\nu^2_{\mathrm{pn}}$ at the VCO then the variogram for the continuous-time PLL is given by \cite{Demir_PN_2006}
\begin{equation}
    \sigma^2_{\theta}(\delta) = \frac{2\pi\beta_{\rm{pn}}}{F_{\mathrm{PLL}}}(1-e^{(-\delta{F_{\mathrm{PLL}}T_{\rm{s}})}}).
\end{equation}

This analysis is then directly used to compute the elements of $\boldsymbol{K}_{\psi}$ for each oscillator type as
\begin{equation}
{K}_{\psi}[m,n] =
\left\{ 
    \begin{array}{lll}
        e^{-2\pi\beta_{\rm{pn}}T_{\rm{s}}|m-n|},&\qquad \mathrm{FRO}  &\\
        e^{-\frac{\pi\beta_{\rm{pn}}}{F_{\mathrm{PLL}}}(1-e^{(-|k-l|{F_{\mathrm{PLL}}T_{\rm{s}})}})}, &\qquad \mathrm{CPLL}
    \end{array} 
\right.
\end{equation}
The full delay-time domain channel estimate is obtained by multiplying the Wiener filtering matrix by the initial snapshot estimate from stage 1,
    $\bG_{\rm DT}^{\mathtt{W}} = \mathbf{W}\widehat{\mathbf{G}}_{\mathrm{TD}}^{\rm{P}}.$
This estimate can be used to perform linear equalization techniques in the delay-time domain techniques after which the equalized data data can be converted to the delay-Doppler domain for detection. It is important to note that the computation of $\mathbf{W}$ is performed offline based on the channel and phase noise statistics. In addition, since it is applied to the overall RIS cascade channel, the complexity of this computation does not increase with the number of RIS elements.

\section{Numerical results}

This section presents numerical results to compare the performance of OTFS to OFDM in RIS-assisted systems using the proposed joint phase noise and channel estimation technique. We also compare the performance of our proposed technique to existing methods which use basis expansion model (BEM) \cite{Sanoop_PCP} and Spline \cite{Thaj_OTSM_2021}. We compare the techniques in terms of normalized mean squared error (NMSE) and BER. Monte Carlo simulation is used to average the results over $10^5$ random channel instances for each simulation. 

\begin{figure}[t]
    \centering
    \includegraphics[width=0.8\linewidth]{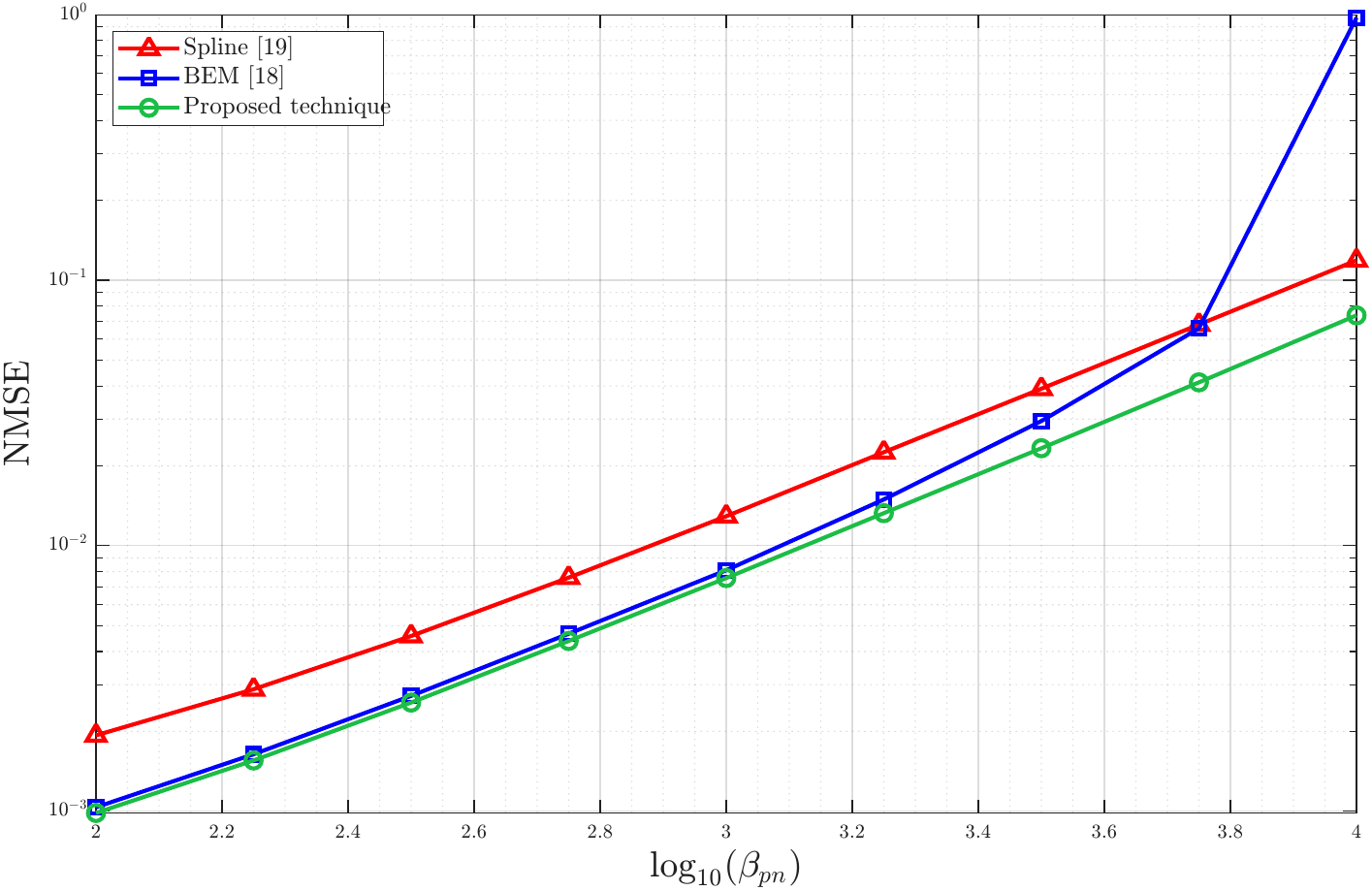}
    \caption{NMSE performance of RIS-OTFS and RIS-OFDM for different phase noise levels.}
    \label{fig:NMSE_PN_only}
\end{figure}

\begin{figure}[t]
    \centering
    \includegraphics[width=0.8\linewidth]{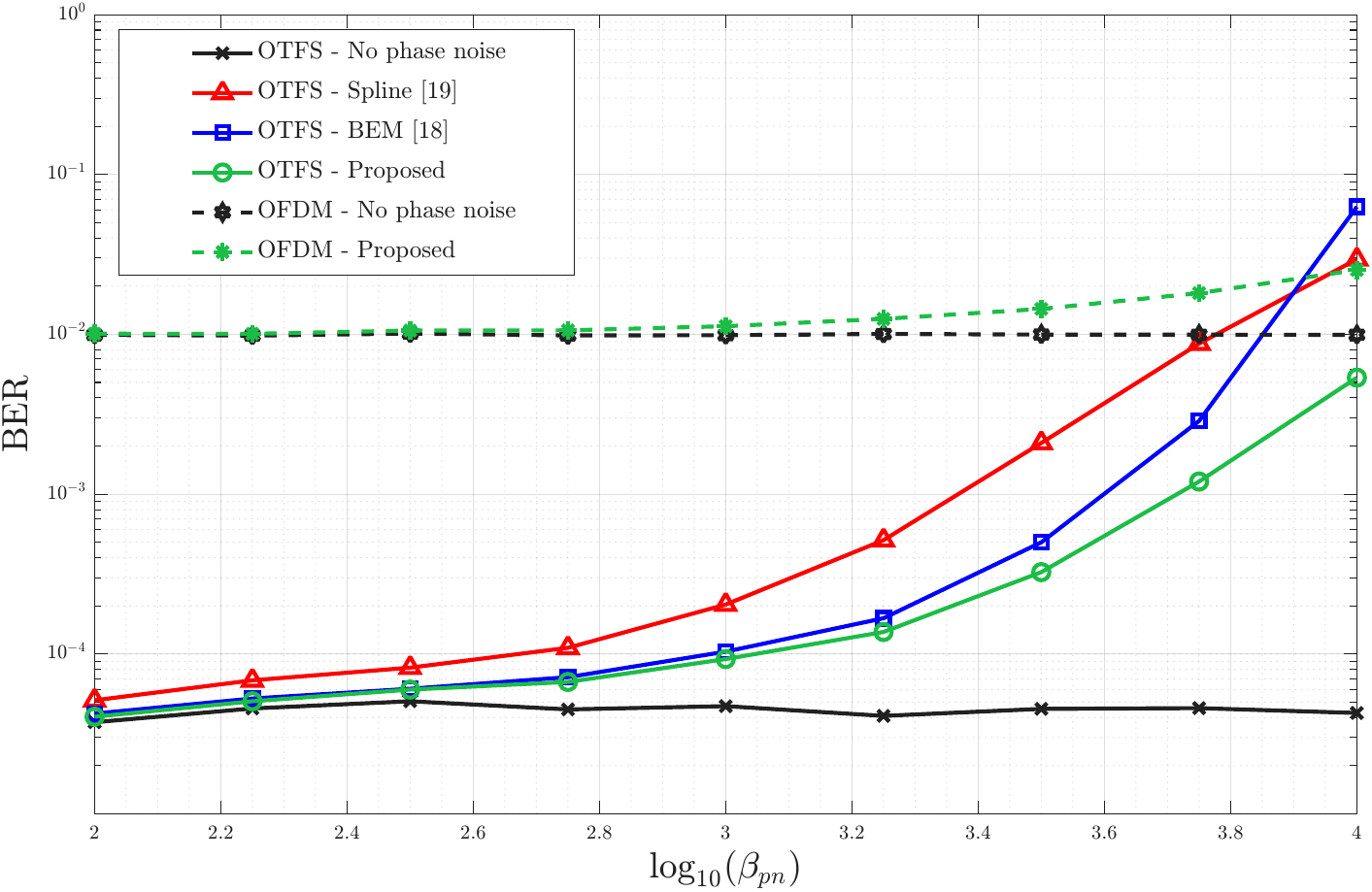}
    \caption{BER performance of RIS-OTFS and RIS-OFDM for 4-QAM for different phase noise levels.}
    \label{fig:BER_PN}
\end{figure}

A carrier frequency of $f_\mathrm{c} = 5.9$ GHz, a transmission bandwidth of 7.68~MHz and a delay-Doppler grid size of $M=128$ and $N=32$ are considered. We consider a planar RIS with 64 elements. The 5G NR Tapped Delay Line C (TDL-C) channel model with a delay spread of 300~ns \cite{3gpp_TS38901} is used to generate the channels for each RIS element. The Doppler shifts are generated using Jakes' model \cite{jakes_model}. For equalization we use the least-squares minimum residual with interference cancellation (LSMR-IC) algorithm \cite{OTFS_LSMR}, with $I_{\rm{ic}} = 5$ and $I_{\rm{lsmr}} = 10$, which correspond to the number of interference cancellation iterations and LSMR iterations, respectively. For BEM, we choose the number of basis functions based on the maximum Doppler spread and the phase noise bandwidth, i.e. $Q_{\rm{BEM}}=\lceil 2k_\mathrm{over}M_{\rm T}N(f_{\rm D}+\beta_{\rm{pn}})T_{\rm s}\rceil+1$, where $k_\mathrm{over}\geq 1$ is the BEM oversampling factor.

In order to focus on the impact of phase noise only, the results shown in Fig. \ref{fig:NMSE_PN_only} concern the scenario where there is no Doppler spread, i.e., $f_{\rm D} = 0$. Fig. \ref{fig:NMSE_PN_only} shows the NMSE as the phase noise bandwidth increases. For this analysis, we consider a fixed SNR level of 0~dB. It can be seen that our proposed method outperforms the existing methods, especially at higher phase noise levels. BEM provides similar performance to proposed method at low phase noise levels but as phase noise increases, BEM faces severe performance degradation. This is because the sample-to-sample channel variations reach a level that BEM cannot capture. Therefore, BEM is ill-suited for estimating phase noise.



The remaining results to be shown concern the scenario where both phase noise and Doppler spread are present. We consider a relative velocity between the transmitter and receiver of 500~km/h in these simulations. Fig. \ref{fig:BER_PN} shows the BER results for an uncoded system where 4-QAM modulation is used as phase noise bandwidth increases. OTFS outperforms OFDM at low to moderate levels of phase noise. However, the BER performance of both degrades as phase noise increases. This indicates that while OTFS does offer improved resilience to PN for RIS-aided systems, the performance can still be heavily impaired by severe phase noise.

\begin{figure}[t]
    \centering
    \includegraphics[width=0.8\linewidth]{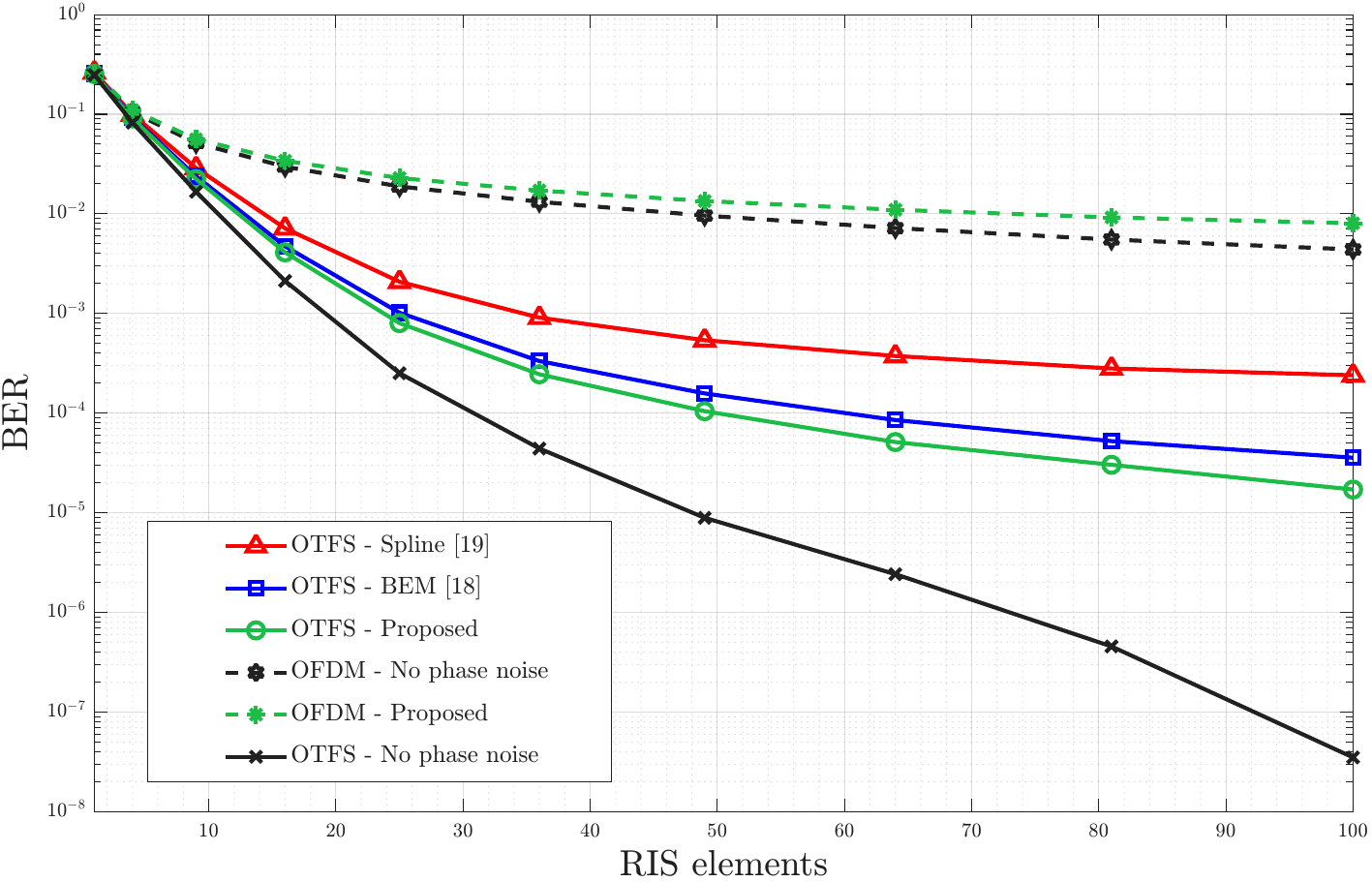}
    \caption{BER performance of RIS-OTFS and RIS-OFDM for 4-QAM for different number of RIS elements.}
    \label{fig:BER_RIS}
\end{figure}

Fig. \ref{fig:BER_RIS} shows the BER performance as the number of RIS elements increases for an uncoded system with 4-QAM modulation and the phase noise bandwidth is set equal to 2~kHz. It can be seen from Fig. \ref{fig:BER_RIS} that RIS-OTFS significantly outperforms RIs-OFDM, providing two orders of magnitude gains. However, there is a loss of performance compared to the case where phase noise is not present. This indicates that while RIS-OTFS provides better performance that RIS-OFDM, phase noise still limits the potential gains of RIS-aided systems. Additionally, our proposed method once again outperforms the BEM and Spline schemes for RIS-OTFS.

\begin{figure}[t]
    \centering
    \includegraphics[width=0.8\linewidth]{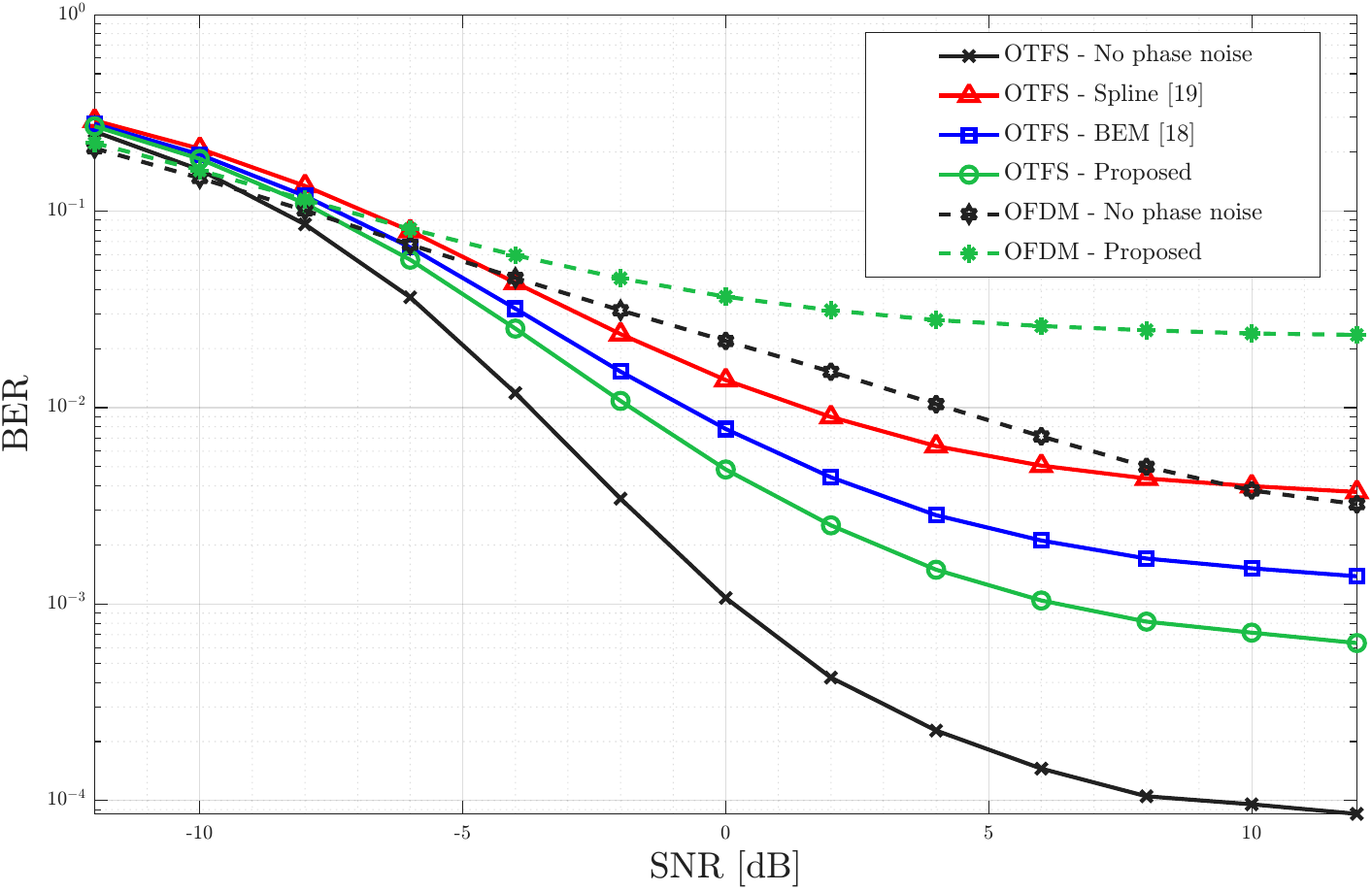}
    \caption{Coded BER performance of RIS-OTFS and RIS-OFDM for 16-QAM at different SNRs.}
    \label{fig:BER_SNR_coded}
\end{figure}

  To make our analysis more realistic, we also study BER performance of a coded system. Fig. \ref{fig:BER_SNR_coded} show the BER results for the coded system where 16-QAM modulation is used. In this analysis, we utilize of a 1/2 rate convolutional encoder. For this simulation, the phase noise bandwidth is set equal to 3~kHz. It can be seen from this figure that RIS-OTFS using the proposed method significantly outperforms RIS-OFDM in both the case with and without phase noise. In fact, RIS-OTFS with the proposed estimation technique offers an order of magnitude improvement over the RIS-OFDM scenario. Additionally, our proposed technique provides approximately 3~dB of gain over the BEM method at a BER of $2\times 10^{-3}$.




\section{Conclusion}
In this paper, we investigated the effect of phase noise on the performance of OTFS and OFDM in RIS-aided systems systems. We proposed a novel technique for joint phase noise and channel estimation for RIS systems using a Wiener filtering based approach. We have presented a range of numerical results which demonstrate that OTFS provides significant gain over OFDM in RIS-aided systems in the presence of phase noise. Our results show the impact of phase noise estimation on the performance of both systems.  Additionally, these numerical results also demonstrate the superiority of our proposed technique over existing methods in terms of BER and NMSE in both coded and uncoded systems.

\bibliographystyle{IEEEtran}
\bibliography{biblio}

\end{document}